\documentstyle[psfig,11pt,a4]{article}

\newcommand{\be}{\begin{equation}}
\newcommand{\ee}{\end{equation}}
\newcommand{\bea}{\begin{eqnarray}}
\newcommand{\eea}{\end{eqnarray}}

\begin{document}

\begin{titlepage}

\begin{flushright}
{\tt
    hep-th/0307096}
 \end{flushright}

\bigskip

\begin{center}

{\bf{\LARGE Generalized Virasoro anomaly \\ for dilaton coupled theories}}

\bigskip
\bigskip\bigskip
 A.  Fabbri$^a$, \footnote{e-mail: \sc
fabbria@bo.infn.it}
S. Farese$^b$ \footnote{\sc farese@ific.uv.es} and
 J. Navarro-Salas$^b$ \footnote{\sc jnavarro@ific.uv.es}

\end{center}

\bigskip%

\footnotesize
\noindent
 a) Dipartimento di Fisica dell'Universit\`a di Bologna and INFN
 sezione di Bologna, Via Irnerio 46,  40126 Bologna, Italy
 \newline
 b) Departamento de F\'{\i}sica Te\'orica and
    IFIC, Centro Mixto Universidad de Valencia-CSIC.
    Facultad de F\'{\i}sica, Universidad de Valencia,
        Burjassot-46100, Valencia, Spain.

\bigskip

\bigskip

\begin{center}
{\bf Abstract}
\end{center}

We derive the anomalous transformation law of the quantum stress tensor
for
a 2D massless scalar field coupled to an external dilaton. This provides a
generalization of the Virasoro anomaly which turns out to be consistent
with the trace anomaly. We apply these results to compute vacuum
polarization of a spherical star based on the equivalence principle.



\end{titlepage}

\newpage

Two-dimensional conformal invariance is a key ingredient to
understand critical behaviour of certain planar statistical
mechanical systems \cite{bpz}. It also plays a pivotal role in the
formulation of superstring theory \cite{polchinski} and in the
quantum mechanics of black holes. The Bekenstein-Hawking area law
is derived in many different ways by applying Cardy's formula for
conformal field theories living in the black hole horizon (see for
instance \cite{car} and references therein). The universal thermal
character of black hole radiation is also related to the fact that
matter fields exhibit two-dimensional conformal invariance in the
vicinity of the
horizon.\\

Many of the basic properties of 2d conformal field theories can be
obtained by studying a simple model, namely a massless scalar
field
 \be \label{mif}
S=-{1\over2}\int d^2x \left(\nabla f\right)^2\>. \ee
Standard canonical quantization and Wick theorem lead to the
well-known operator product expansion of the quantum stress tensor
\bea T_{\pm\pm}(x^{\pm}) T_{\pm\pm}( x'^{\pm}) &=& \frac{1}{8\pi
^2 (x^{\pm}- x'^{\pm})^4}- \frac{1}{\pi (x^{\pm}- x'^{\pm})^2}
T_{\pm\pm}( x'^{\pm}) \nonumber \\ &-&
\frac{1}{2\pi (x^{\pm}- x'^{\pm})}\partial_{\pm}
T_{\pm\pm}( x'^{\pm}) +\ ... \ , \eea
where $x^{\pm}=x^0 \pm x^1$ are null Minkowskian
coordinates. The above expansion leads to the Lie algebra \bea
\left[  T_{\pm\pm}(x^{\pm}),  T_{\pm\pm}(x'^{\pm }) \right] &=&
\frac{1}{2\pi} \partial_{x^{\pm}}\delta(x^{\pm}-x'^{\pm })
T_{\pm\pm}(x'^{\pm }) -\frac{1}{96\pi^2}
\partial_{x^{\pm}}^3 \delta (x^{\pm} - x'^{\pm }) \nonumber \\ &-&
\left( x^{\pm}\leftrightarrow x'^{\pm} \right)
\ .\eea
Since  $T_{\pm\pm}(x^{\pm})$, up to normalization, are the generators of
infinitesimal conformal transformations $x^{\pm} \to x^{\pm}+ \epsilon
^{\pm}(x^{\pm})$, this implies the following infinitesimal transformation law
for the stress tensor  \be \label{actio}\delta_{\epsilon ^{\pm}}
T_{\pm\pm}=\epsilon ^{\pm}
\partial_{\pm} T_{\pm\pm} +2\partial_{\pm}\epsilon ^{\pm}
T_{\pm\pm}-\frac{1}{24\pi}\partial_{\pm}^3\epsilon ^{\pm}\ .\ee
Exponentiating the action (\ref{actio}) one gets, under the conformal
transformation $x^{\pm} \to y^{\pm}(x^{\pm})$, the following
anomalous transformation law \be \label{trl}  T_{\pm\pm}(y^{\pm})
= \left( \frac{dx^{\pm}}{dy^{\pm}}\right)^2  T_{\pm\pm}(x^{\pm}) -
\frac{1}{24\pi}\{ x^{\pm}, y^{\pm} \} \ ,\ee where $\{ x^{\pm}, y^{\pm}\}=
\frac{\partial^3 x^{\pm}}{\partial y^{\pm 3}}/ \frac{\partial x^{\pm
}}{\partial y^{\pm}} -\frac{3}{2} \left(\frac{\partial^2 x^{\pm}}{\partial
y^{\pm 2}}/\frac{\partial x^{\pm }}{\partial y^{\pm}}\right)^2$ is the
Schwarzian derivative. All these expressions can be regarded as
different realizations of the so-called Virasoro anomaly. For a
generic conformal field theory the above results are valid
provided we multiply the $c$-number terms of the above equations
by the central charge $c$ characterizing the
particular theory \cite{bpz}. \\

The aim of this work is to study the modification of the transformation law
(\ref{trl}), when a dilaton field $\phi$ is present and
(\ref{mif}) is replaced by \be \label{mifcd} S=-{1\over2}\int
d^2x  e^{-2\phi}\left(\nabla f\right)^2\>. \ee A nice
justification of the form of the dilaton coupling comes from General
Relativity. If a scalar field $f$ is minimally coupled to a 4D
spherically symmetric metric \be
ds^{2}_{(4)}=ds^{2}_{(2)}+e^{-2\phi}d\Omega^2\ , \ee and we perform
dimensional reduction from $-\frac{1}{8\pi }\int d^4
x\sqrt{-g}(\nabla f)^2, $
we obtain the above action (\ref{mifcd}) in case of flat 2d space. \\

Let us now consider a simple case, namely the one associated to
the four-dimensional Minkowski space. In this situation it is
$ds^{2}_{(2)}=-dx^+dx^-$, where $x^{\pm}=t\pm r$, and
$e^{-2\phi}=r^2$. The mode expansion of the field $f$ living in
the $t-r$ plane (with the condition $f(r=0)=0$) is
\be f=\int_{0}^{\infty} \frac{dw}{\sqrt{4\pi
w}}\left[ a_w (e^{-iwx^+} - e^{-iwx^-}) + a_w^\dagger (e^{iwx^+} -
e^{iwx^-}) \right] e^{\phi}\ .\ee The null components of the
stress tensor are given by \be
T_{\pm\pm}(x^+,x^-)=e^{-2\phi}(\partial_{\pm} f)^2, \ee and the
corresponding normal ordered operators can be defined, as usual,
by point-splitting \be T_{\pm\pm}(x^+,x^-) = \lim_{x^{\pm}\to x'^{\pm }}
e^{-(\phi(x) +\phi(x'))} \frac{\partial}{\partial
x^{\pm}}\frac{\partial}{\partial x'^{\pm }} (f(x)f(x')- \left<
f(x)f(x')\right>), \ee where the two-point function is \be \left<
f(x)f(x') \right> = -\frac{1}{4\pi} e^{\phi(x)+\phi(x')}\ln\frac{
(x^{+ }-x'^{+})(x^- - x'^{-})}{(x^{+ }-x'^{-})(x^- - x'^{+})}\
.\ee Under a conformal transformation $x^{\pm}\to y^{\pm}(x^{\pm})$ normal
ordering breaks covariance and the transformed stress tensor picks
up the following anomalous non-tensorial contributions \bea
\label{trlm}
 T_{\pm\pm}(y^+,y^-) &=& \left( \frac{dx^{\pm}}{dy^{\pm}}\right)^2
 T_{\pm\pm}(x^+,x^-) -\frac{1}{24\pi}\{ x^{\pm}, y^{\pm} \}\nonumber \\
&-& \frac{1}{4\pi} \left[ \frac{d^2 x^{\pm}}{dy^{\pm
2}} \left(\frac{dx^{\pm}}{dy^{\pm}} \right)^{-1}
\frac{\partial\phi}{\partial y^{\pm}} + \ln \left( \frac{dx^+}{dy^+}
\frac{dx^-}{dy^-} \right) 
\left( \frac{\partial\phi}{\partial y^{\pm}}\right)^2 \right] \ . \eea

This expression generalizes the Virasoro-type transformation law
(\ref{trl}) by adding terms depending on the derivatives of
$\phi$. At this point we would like to remark that the above
expression has been obtained for a particular form of $\phi$ in
terms of the null coordinates $x^{\pm}$, namely \be \label{fldi}
\phi=-\ln \frac{x^+ - x^-}{2} .\ee However we want to stress that
the result has general validity, irrespective of the particular
form of the external dilaton field. We shall prove
this in two different ways: \\
i) the short-distance behaviour for the Hadamard function  does not depend on the
specific model; \\
ii) we shall show that eq. (\ref{trlm}) is the only local expression which is
consistent with the trace anomaly derived in the context of
gravitational physics \cite{mukanov-wipf-zelnikov}.
\\

We point out that the conformal symmetry can be recovered in regions where
$\partial_{\pm}\phi \to 0$. This happens typically when $r$ approaches infinity
and also, in the context of curved spacetime, at the black hole horizons.
\\

The equation of motion for the field $f$, derived from the action
(\ref{mifcd}), is \be\label{eom} \partial_+(e^{-2\phi}\partial_-
f)+\partial_-(e^{-2\phi}\partial_+ f)=0. \ee In general this can
be solved only for particular forms of $\phi$, for instance in the
situation where it is given by (\ref{fldi}) or \be \phi =
-\frac{1}{2}\ln\frac{(x^+ - x^-)}{2}\ .\ee In the latter case the
equation of motion for $f$ (\ref{eom}) coincides with the equation of
a minimal scalar
field in a three-dimensional spacetime, described by the action
\be S=-\frac{1}{4\pi}\int d^3 x \sqrt{-g}(\nabla f)^2\ , \ee
under the assumption of
axi-symmetry for the field $f$ and the metric
$ds^2_{(3)}=ds^2_{(2)}+r^2 d\varphi^2 $,
  where the radial function is given by $r=e^{-2\phi}$.
This equation turns out to be equivalent to one equation of the
Einstein-Rosen subsector of pure General Relativity. The system
is exactly solvable both classically and quantum-mechanically
(details can be found in \cite{Kuchar},
\cite{Ashtekar-Angullo-Cruz}, \cite{Niedermaier}, \cite{Barbero})
and, therefore, it can provide a nontrivial test of the formula
(\ref{trlm}). The field $f$ can be expanded in modes as follows
\be f=\int_{0}^{\infty} \frac{dw}{\sqrt{2}} J_0(r w)\left[ a_w
e^{-iwt} +a^\dagger_w e^{iwt}\right] \ee where $J_0$ is the zero
order Bessel function. At the quantum level the coefficients $a_w$
and $a_w^\dagger$ are converted into annihilation and creation
operators obeying the commutation relation
$[a_w,a_{w'}^\dagger]=\delta(w-w')$. To work out the quantum
behaviour of the stress tensor we need to evaluate the Hadamard
function $G^{(1)}(x,x')\equiv \frac{1}{2}\left< 0|\{ f(x), f(x')\}
|0\right>$. This turns out to be equal to \cite{Barbero2},
\cite{Niedermaier}\\
i) for $0<|t'-t|<|r'-r|$

$$G^{(1)}(x,x')=\frac{1}{\pi\!\sqrt{[(r'\!+\!
\!r)^2-(t'\!-\!\!t)^2]}}
K\!\!\left(\!\!\sqrt{\!\frac{4rr'}{(r'\!+\!
\!r)^2-(t'\!-\!\!t)^2}}\right);$$ \\
ii) for $|r'-r|<|t'-t|<r'+r $
$$
G^{(1)}(x,x')=\frac{1}{2\pi}\frac{1}{\sqrt{ r
r'}}\,K\left(\sqrt{\frac{(r'\!+\!r)^2-(t'\!-\!t)^2}
{4rr'}}\right).
$$
iii) for $r+r'<|t'-t|$ it is $G^{(1)}(x,x')=0$,\\
where $K(k)=\int_0^{\pi/2} d\theta /\sqrt{1-k^2\sin^2(\theta)}$ is
the complete elliptic integral. Using the expansion \cite{GR} 
\be K(k')= \ln\frac{4}{k'} + (\frac{1}{2})^2 \left(
\ln\frac{4}{k'} - 1\right) k'^{2} +O(k'^4 \ln\frac{4}{k'}), \ee where
$k'=\sqrt{1-k^2}$, we obtain \bea\label{haex}
G^{(1)}(x,x')&=&-\frac{e^{\phi(x)+\phi(x')}}{4\pi} [ \ln(x^{
+}-x'^+)(x^{-}-x'^-) + const. \nonumber \\ &+&
 O\left(  (x^{+}-x'^+)(x^{-}-x'^-)\ln(x^{+}-x'^+)(x^{-}-x'^-)\right) ]\ .\eea
In the computation of the transformation law of the stress tensor,
via point-splitting, only the leading term in (\ref{haex})
produces a nontrivial contribution. Therefore it is easy to see
that the final result is (\ref{trlm}). Moreover, the above
expression agrees with the De Witt-Schwinger expansion of
$G^{(1)}(x,x')$, restricted to flat space-time, given in
\cite{bunch-christensen-fulling},
 \cite{balbinot-fabbri-nicolini-frolov}
\be G^{(1)}(x,x')=\frac{e^{\phi(x)+\phi(x')}}{2\pi}\left[ -(\gamma
+\frac{1}{2}\ln\frac{m^2\sigma}{2}) +O(\sigma\ln\sigma) \right]\ ,\ee
where $\gamma$ is the Euler constant,  $m^2$ is an infrared cutoff
and  $\sigma$ is one half the square of the distance between the
points $x$ and $x'$.\\

Due to presence of $\phi$ the classical conservation laws $\partial_{\mp}
T_{\pm\pm}=0$ get modified to (see \cite{balbinot-fabbri}, \cite{kummer-vassilevich}
for a
higher-dimensional interpretation) \be \label{qucl}
\partial_{\mp} T_{\pm\pm} + \partial_{\pm}\phi \frac{\delta S}{\delta \phi}=0,\ee
where \be \frac{\delta S}{\delta\phi}=
-2e^{-2\phi}\partial_+ f\partial_- f\ .\ee Let us analyze the
quantum analogous of these equations. The transformation
law for $\left<  T_{\pm\pm} \right>$ is given by eq. (\ref{trlm}) and
the corresponding one for $\left< \frac{\delta S}{\delta \phi}
\right>$ should be, on general grounds, of the form \be
\label{prmo} \left< \frac{\delta S}{\delta \phi} (y^{\pm}) \right> =
\frac{dx^+}{dy^+}\frac{dx^-}{dy^-} \left<  \frac{\delta S}{\delta
\phi} (x^{\pm}) \right> +  \Delta (\phi ;x^{\pm},y^{\pm}).\ee Let us suppose
that \be \label{clqu}
\partial_{\mp}\left< T_{\pm\pm}\right> +
\partial_{\pm}\phi  \left< \frac{\delta S}{\delta 
\phi}\right>=0.\ee
If we transform this relation according to (\ref{trlm}) and
(\ref{prmo}) we get, by consistency, \bea &-& \frac{1}{4\pi}\frac{
\frac{\partial^2 x^{\pm}}{\partial y^{\pm 2}}}{ \frac{\partial
x^{\pm}}{\partial y^{\pm}}} \frac{\partial}{\partial
y^+}\frac{\partial}{\partial y^-}\phi -\frac{1}{2\pi}\ln
\left( \frac{\partial x^+}{\partial y^+}  \frac{\partial x^-}{\partial
y^-}\right)  (\frac{ \partial\phi}{\partial y^{\pm}})\frac{\partial}{\partial
y^+} \frac{\partial}{\partial y^-}\phi \nonumber \\ &-&
\frac{1}{4\pi} \left( \frac{\partial \phi}{\partial y^{\pm}}\right)^2
\frac{ \frac{\partial^2 x^{\mp}}{\partial y^{\mp 2}}}{ \frac{\partial
x^{\mp}}{\partial y^{\mp}}} + \frac{\partial \phi}{\partial 
y^{\pm}} \Delta
(\phi ;x^{\pm},y^{\pm}) =0 .\eea These two equations are compatible with
the uniqueness of $\Delta(\phi ;x^{\pm},y^{\pm})$ only if \be \label{requ}
\Box\phi=(\nabla\phi)^2.\ee If $\phi$ does not obey (\ref{requ})
the quantum transformation law (\ref{clqu}) must be modified. We
find that the only possibility to maintain consistency with the
transformation law (\ref{trlm}) is by adding a nontrivial trace
$\left< T_{+-}\right>$ just of the form \be \label{fsan} \left<
T_{+-}\right> = -\frac{1}{4\pi}\left(
\partial_+\phi \partial_-\phi -
\partial_+\partial_-\phi \right).\ee
Finally, the quantum conservation law, invariant under conformal
transformations, reads \be \label{qcl}
\partial_{\mp}\left< T_{\pm\pm}\right> + \partial_{\pm} \left< T_{+-}\right>
+\partial_{\pm}\phi \left< \frac{\delta S}{\delta \phi}\right> =0. \ee
We have to point out that the anomalous trace derived in this
approach agrees with the one derived in curved space-time
(\cite{mukanov-wipf-zelnikov}, see also \cite{molti})  . For the dilaton-coupled
theory the trace anomaly, obtained in a covariant quantization
scheme, is \be \label{tranom} \left< T \right> = \frac{1}{24\pi}\left( R -
6(\nabla\phi)^2 + 6\Box\phi \right) .\ee If we restrict to flat
space-time we obtain (\ref{fsan}). So our derivation can be seen,
as a by-product, as an alternative and simple way to get the
dilaton contribution to the trace anomaly. Moreover, the argument
can be applied the other way around: assuming (\ref{fsan}),
(\ref{qcl}) and locality one gets the $\phi$ dependent terms of (\ref{trlm}).\\

Finally, we want to explain briefly an application of
our results to gravitational physics. We shall compute the vacuum
polarization of a spherical star using the anomalous
transformation law (\ref{trlm}) and the help of the equivalence principle
to deal with curved space. We 
consider the Schwarzschild spacetime,
for which the 2d metric is \be
ds^2_{(2)}=-(1-2M/r)dudv\ , \ee
where $u$ and $v$
are the Eddington-Finkelstein null coordinates $v=t+r^*$,$u=t-r^*$,
$r^*=r+2Mln(\frac{r}{2M}-1)$, and
the dilaton is \be \label{sdilaton}
e^{-2\phi}= r(u,v)^2\ . \ee In a generic point $X$ of the
four-dimensional space-time one can always introduce locally
inertial coordinates $\xi_{X}^{\alpha}$. Restricting now our
attention to the $(t,r^*)$-sector we can then construct the
corresponding null coordinates $\xi_{X}^{\pm}$ and the local vacuum
state  $|0_{X}>$. We shall make the natural assumption
that for the state $|0_{X}>$  we have \be
\label{evlv} <0_{X}|T_{\pm \pm}(\xi_{X}^{\pm}(X))|0_{X}>=0. \ee Now
we want to work out the same quantities with respect to the
(global) vacuum state defined with respect to the modes which asymptotically
behave as
$\frac{e^{-i\omega u}}{r}$, $\frac{e^{-i\omega v}}{r}$.
This is the so-called Boulware
state $|B>$ \cite{boulware}. We can obtain $\left< B| T_{\pm\pm}(X)|B \right>$
 by using the anomalous transformation law (\ref{trlm})
 between $\{ \xi_X^{\pm} \}$ and $\{ x^+=v, x^-=u \}$.
 Up to second order and Poincar\'e
 transformations these transformations are unambiguous
  and can be chosen to be conformal \cite{weinberg}
 \be
 \xi^{\pm}_X = b^{\pm}_{\pm}\left[ (x^{\pm} - x^{\pm}(X)) +
 \frac{\Gamma^{\pm}_{\pm\pm}}{2}(x^{\pm} - x^{\pm}(X))^2
 + F_{\pm}(x^{\pm}-x^{\pm}(X))^3  + ... \right] \ .\ee
 In a conformal frame $ds^2=-e^{2\rho}dx^+dx^-$
 the constants $b^{\pm}_{\pm}$ satisfy the constraint $b^+_+b^-_-=e^{2\rho(X)}$
 and $\Gamma^{\pm}_{\pm\pm}=2\partial_{\pm}\rho$.
 Note that the Schwarzian derivative requires the third order as well,
 which is not determined by the requirement that $\xi^{\pm}_X$ are locally inertial.
 We naturally fix it by imposing that, for a flat
 metric, $\xi^{\pm}(X)$ are the global null minkowskian coordinates. This leads to
 \be
 F_{\pm}=  \frac{1}{3}\partial_{\pm}\rho(X) + \frac{2}{3}\left(
 \partial_{\pm}\rho(X)\right)^2  \ .\ee
 The result is
 \bea \left< B|T_{\pm\pm}(x^{\pm}(X))|B\right> &=& -\frac{1}{12\pi}
 \left( (\partial_{\pm}\rho)^2 - \partial_{\pm}^2\rho \right)(X) + \frac{1}{2\pi}
 \partial_{\pm}\rho(X)\partial_{\pm}\phi (X) \nonumber \\
 &+&\frac{1}{2\pi}\rho(X)(\partial_{\pm}\phi(X))^2 \ .\eea
We remark that neglecting the terms containing the dilaton these are the null
components of the stress tensor derived from the Polyakov effective action \cite{polyakov}.
We finally obtain, for a generic point,
\be \label{boulsta}
\left< B| T_{\pm\pm}|B\right> = \frac{1}{24\pi}\left( -\frac{4M}{r^3}+
\frac{15}{2}\frac{M^2}{r^4}\right) + \frac{1}{16\pi r^2} (1-\frac{2M}{r})^2 
\ln (1-\frac{2M}{r}) \ .
\ee
The $+-$ component is state independent and fixed by the trace anomaly (\ref{tranom})
\be \label{boulsta2}
\left< B|T_{+-}|B\right>= \frac{1}{12\pi}(1-\frac{2M}{r})\frac{M}{r^3}\ .\ee
Similar results, based on exact properties of the effective action under 
Weyl transformations, 
were derived in \cite{balbinot-fabbri2}. 
 These expressions correctly vanish in the
limit $M\to 0$ as well as for $r\to \infty$. Moreover, in the horizon limit they
are in agreement
with the results derived from canonical quantization 
\cite{balbinot-fabbri-nicolini-frolov}.

To end the paper, we would like to remark that the fact that
(\ref{trlm})  is the exact transformation law of
the quantum
stress tensor for a generic dilaton field $\phi$ should not be a surprise
at all. One of
the main features of 2d conformal field theories is the existence of
universal behaviours,
irrespective of the particular model considered. Therefore one could be
tempted to conjecture
that (\ref{trlm}) is also valid for an arbitrary conformal field theory coupled to a
dilaton, up
to numerical coefficients in the $c$-numer terms.

 \section*{Acknowledgements}
This research has been partially supported by the research grants
BFM2002-04031-C02-01 and BFM2002-03681 from the Ministerio de Ciencia y
Tecnologia (Spain), EU FEDER funds and the INFN-CYCIT
Collaborative Program. S.F. acknowledges
the Ministerio de Educacion, Cultura y Deporte
for a FPU fellowship. J. N-S would like to acknowledge the Department of Physics
of the University of Bologna for hospitality
A.F. and J. N-S thank R. Balbinot and S. Fagnocchi for useful discussions.
Finally, we wish to thank the authors of \cite{Barbero} 
for assistance concerning the 2+1 model considered.

\end{document}